\documentclass{article}

\usepackage{amsfonts}

\title{A Completely Invariant SUSY Transform of Supersymmetric QED}
\author{M.L. Walker}
\date{}

\begin{document}

\maketitle

\begin{abstract}
We study the SUSY breaking of the covariant gauge-fixing term in SUSY QED
and observe that this corresponds to a breaking of the Lorentz
gauge condition by SUSY. Reasoning by analogy with SUSY's violation of the 
Wess-Zumino gauge, we argue that the SUSY transformation, already modified to
preserve Wess-Zumino gauge, should be further modified by another gauge
transformation which restores the Lorentz gauge condition. We derive this  
modification and use the resulting transformation to derive
a Ward identitiy relating the photon and photino propagators without using
ghost fields. Our transformation also
fulfills the SUSY algebra, modulo terms that vanish in Lorentz gauge.
\end{abstract}

\section{Introduction} \label{sec:intro}
Not only is supersymmetry (SUSY) believed necessary for the non-trivial
unification of gravity and the gauge forces, but it is also expected to 
manifest in physics not much higher than the electroweak scale. 
One particular topic of current interest~\cite{RSS01,OR02} 
is the effect of the covariant gauge-fixing
term in the component formalism. 
This term is not invariant to SUSY transformations
in the component foralism and the non-linear corrections of Wess and Zumino
do not restore it.

The conventional method of handling this situation is to introduce ghost fields
and find SUSY BRST transformations~\cite{RSS01,OR02,PS86}. 
Ghost fields are long known to be 
an essential part of non-Abelian field theories~\cite{IZ80}, 
and their use in SUSY theories
for linearising the transformations and removing the auxiliary fields off 
mass-shell is also well-established~\cite{PS86}. 
The former of these objectives is
considered necessary for the derivation of SUSY Slavnov-Taylory identities
to relate the vertices of SUSY gauge theories~\cite{PS86}, although the 
decoupling of the ghost fields in the Abelian
case allows one to assume the invariance of the effective action even under
the non-linear Wess-Zumino transformations and derive SUSY Ward 
identities~\cite{WB99b}. The latter objective arises because, while 
it is understandably desirable to remove the auxiliary fields, to do so 
off-mass shell introduces errors that must be compensated.
An alternative to this approach, in which the auxiliary fields are not
removed but the scalar fields are instead ``reinterpreted'', 
has also been demonstrated~\cite{WB99b} but is not in common use. 

While we do not dispute these particular applications, we do dispute that
the introduction of ghosts is necessary to obtain transformations that leave 
the entire SUSY QED action invariant. 
The conventional view seems to be that since
the SUSY transformations found by Wess and Zumino in the seminal 
paper in which they constructed SUSY QED only apply to the
gauge-invariant part of SUSY QED~\cite{WZ74}, then the 
gauge-fixing spoils, or at least limits,
SUSY invariance. We argue instead that these transformations are 
incomplete, and
then derive their completion. This turns out to be nothing more than the
addition of another non-linear gauge transformation, similar to those
found in Wess and Zumino's original paper.

A highly technical calculation by \cite{RSS01} finding the Noether current for
the conventional SUSY BRST transformations in the
component formalism, finds that SUSY is only a symmetry of the on-shell
states, but not of the entire Fock space. In Lorentz gauge
they found that SUSY is violated when the
fields move off mass-shell by an amount proportional to
\begin{equation} \label{eq:SUSYbreak}
\delta_S (\partial \cdot A) = \bar{\zeta} \not \! \partial \lambda,
\end{equation}
where $A_\mu,\lambda$ are the gauge and gaugino field respectively
and $\zeta$ is the SUSY transform 
parameter. As the authors correctly assert, the discrepency arises from the
SUSY breaking of the gauge-fixing term.

A later work~\cite{OR02} 
calculates the SUSY Ward identities for the Green's
functions in SUSY gauge theories and finds 
that they do not hold at tree level, even
in the Abelian case. The authors make 
the erroneous claim that the discrepency vanishes on mass-shell, and construct
BRST identities whose corresponding Slavnov-Taylor identities do hold at tree
level.

In this paper we go back to where Wess and Zumino left off and reexamine
the SUSY breaking of the gauge-fixing term. In sec.~\ref{sec:WZL} we derive
the alteration to the conventional SUSY transformations needed
to leave the covariant gauge-fixing term invariant also, and find that the
resulting transformation obeys the SUSY algebra in Lorentz gauge. 
In sec.~\ref{sec:SWI} we use our newfound transformation to derive a
SUSY Ward identity relating the photon and photino propagators, and another
relating the electron and selectron propagators. We conclude that the 
gauge-fixing term presents neither a fundamental difficulty, nor any 
unintuitive alterations to the Green's functions of SUSY QED. Instead, we find 
that SUSY Ward identities \textit{can} be found, and that the claims 
of~\cite{RSS01,OR02} are artifacts of working with BRST identities based 
on incomplete SUSY transformations.

\section{The Completely Invariant SUSY Transformations} \label{sec:WZL}

As is well known~\cite{W90}, 
the matter fields in SUSY QED form a chiral multiplet
\begin{eqnarray} \label{eq:chiral}
\delta_S a &=& -i\bar{\zeta} \psi \nonumber \\
\delta_S b &=& \bar{\zeta} \gamma_5 \psi \nonumber \\
\delta_S \psi &=& (f+i\gamma_5 g)\zeta 
+ i\not \! \partial (a+i\gamma_5 b) \zeta \nonumber \\
\delta_S f &=& \bar{\zeta} \not\!\partial \psi \nonumber \\
\delta_S g &=& i\bar{\zeta} \gamma_5 \not\!\partial \psi, 
\end{eqnarray}
while the gauge field is part of a more general multiplet,

\begin{eqnarray} \label{eq:general}
\delta_S C & = & \bar{\zeta} \gamma_5 \chi \nonumber \\
\delta_S \chi & = & (M + i\gamma_5 N ) \zeta + i\gamma^\mu (A_\mu + i\gamma_5
\partial_\mu C) \zeta \nonumber \\
\delta_S M & = & \bar{\zeta} ( \not\!\partial \chi + i\lambda ) \nonumber \\
\delta_S N & = & i\bar{\zeta} \gamma_5 ( \not\!\partial \chi + i\lambda ) 
\nonumber \\
\delta_S A_\mu & = & \bar{\zeta} \gamma_\mu \lambda - i\bar{\zeta} \partial_\mu
\chi \nonumber \\
\delta_S \lambda & = & \frac{1}{2} 
(\gamma^\nu \gamma^\mu - \gamma^\mu \gamma^\nu )
\partial_\mu A_\nu \zeta + i\gamma_5 D \zeta \nonumber \\
\delta_S D & = & i\bar{\zeta} \gamma_5 \! \not\!\partial \lambda.
\end{eqnarray}

The elements $C$ through to $N$ are gauge degrees of freedom and it is well
known that they are easily removed with a gauge transformation~\cite{WZ74,W90}.
The multiplets are combined in a gauge independant way to give the Lagrangian
\begin{eqnarray} \label{eq:lagrang}
L &=& |f|^2 + |g|^2+ |\partial_\mu a|^2 + |\partial_\mu b|^2
- \bar{\psi} \not \! \partial \psi \nonumber \\ \nonumber \\
&&-m(a^*f + af^* + b^*g + bg^* + i\bar{\psi} \psi)\nonumber \\ \nonumber \\
&&- ieA^\mu(a^\ast \stackrel{\leftrightarrow}{\partial}_\mu a
+ b^\ast \stackrel{\leftrightarrow}{\partial}_\mu b
+ \bar{\psi} \gamma_\mu \psi) \nonumber \\ \nonumber \\
&&- e[\bar{\lambda}(a^\ast + i\gamma_5 b^\ast)\psi 
- \bar{\psi}(a + i\gamma_5 b) \lambda] \nonumber \\ \nonumber \\
&&+ ieD(a^\ast b - a b^\ast)
+e^2 A_\mu A^\mu (|a|^2 + |b|^2) \nonumber \\ \nonumber \\
&&-\frac{1}{4}F^{\mu \nu}F_{\mu \nu} - 
\frac{1}{2}\bar{\lambda} \not \! \partial \lambda + \frac{1}{2} D^2
- \frac{1}{2\xi}(\partial_\mu A^\mu)^2 ,
\end{eqnarray}
where the gauge dependant superpartners of $A_\mu,\lambda$ and $D$ have 
been gauged away (Wess-Zumino gauge).

As already noted, the gauge-fixing term spoils the SUSY invariance of the 
action. This is similar to the situation encountered by Wess and Zumino 
in their construction of the gauge invariant part of
SUSY QED~\cite{WZ74}. Rather than invoke ghost 
fields, they instead realised that the violation is due to the spoiling of
Wess-Zumino gauge by SUSY tranformations. Their remedy was to
follow the original SUSY transformation with a gauge transformation
that restored their gauge. 

It is the same here. The Lorentz gauge sets the longitudinal part of the gauge
field to zero but the SUSY transformation contributes to the longitudinal
part so that it is no longer zero. The original 
SUSY transformation spoils the Lorentz gauge, just as it spoils the 
Wess-Zumino gauge. The remedy is the same: follow the SUSY transformation
with a gauge transformation to restore the gauge. Such a transformation must
exist. To find it, observe that we are looking for a gauge parameter 
$\theta$ such that
\begin{equation}
\partial^2 \theta_a = \delta_S (\partial \cdot A).
\end{equation}
From this and eqn~(\ref{eq:SUSYbreak}) it follows that
\begin{equation}
\theta_a = -\bar{\zeta} \frac{\not \! \partial}{\Box} \lambda.
\end{equation}

The invariant SUSY transformation for SUSY gauge 
theories in component form and Lorentz gauge, or the Wess-Zumino-Lorentz gauge,
is $\delta_{WZL} = \delta_{S} + \delta_{WZ} + \delta{L}$ where $\delta_S$ is
the original SUSY transformation 
given by eqs.~(\ref{eq:chiral},\ref{eq:general}) in Wess-Zumino gauge while
$\delta_{WZ}$ is given by
\begin{eqnarray} \label{eq:deltaWZ}
\delta_{WZ} a &=& 0, \nonumber \\
\delta_{WZ} b &=& 0, \nonumber \\
\delta_{WZ} \psi &=& -e \not \! A (a-i\gamma_5 b)\zeta, \nonumber \\
\delta_{WZ} f &=& -e\bar{\zeta} [a\lambda +i b \gamma_5 \lambda 
-i\not \! A \psi ], \nonumber \\
\delta_{WZ} g &=& -ei\bar{\zeta} [\gamma_5 \lambda +i b \lambda 
+i \not \! A \gamma_5 \psi], \nonumber \\
\delta_{WZ} A_\mu &=& 0, \nonumber \\
\delta_{WZ} \lambda &=& 0, \nonumber \\
\delta_{WZ} D &=& 0, 
\end{eqnarray}
and $\delta_L$ by
\begin{eqnarray} \label{super}
\delta_L a &=& i\bar{\zeta} \frac{\not \! \partial}{\Box} \lambda a, 
\nonumber \\
\delta_L b &=& i\bar{\zeta} \frac{\not \! \partial}{\Box} \lambda b, 
\nonumber \\
\delta_L \psi &=& i\bar{\zeta} \frac{\not \! \partial}{\Box} \lambda \psi, 
\nonumber \\
\delta_L f &=& i\bar{\zeta} \frac{\not \!\! \partial}{\Box} \lambda f, 
\nonumber \\
\delta_L g &=& i\bar{\zeta} \frac{\not \!\! \partial}{\Box} \lambda g, 
\nonumber \\
\delta_L A_\mu &=& \partial_\mu 
\bar{\zeta} \frac{\not \! \partial}{\Box} \lambda , \nonumber \\
\delta_L \lambda &=& 0, \nonumber \\
\delta_L D &=& 0, 
\end{eqnarray}

That $\delta_{WZL}$ leaves the action invariant is obvious. The gauge-invariant
part, constructed by Wess and Zumino~\cite{WZ74}, was shown by them to be
invariant under both $\delta_{S} + \delta_{WZ}$ and any standard Abelian 
gauge transformation, including $\delta_L$. The gauge-fixing term is unaffected
by $\delta_{WZ}$ while $\theta$ was chosen so that 
$\delta_S + \delta_L$ would leave it invariant.

Less obvious, though not surprising, is that $\delta_{WZL}$ obeys
the SUSY algebra in Lorentz gauge. For example, 
\begin{equation} \label{eq:Asusyalg}
[\delta_{WZL1}, \delta_{WZL2}] A_\mu 
= \bar{\zeta}_2 \not \! \partial \zeta_1 A_\mu 
+i \bar{\zeta}_2 \not \! \partial \zeta_1 \frac{\partial_\mu}{\Box}
\partial \cdot A
\end{equation}
where the final term vanishes in Lorentz gauge. Similarly for the
electron field
\begin{equation} \label{eq:psisusyalg}
[\delta_{WZL1}, \delta_{WZL2}] \psi
=\bar{\zeta}_2 \not \! \partial \zeta_1 \psi
-i \bar{\zeta}_2 \frac{\not \! \partial}{\Box} \zeta_1 \partial \cdot A \psi.
\end{equation}
Similar results hold for the other fields.

\section{SUSY Propagator Ward Identities} \label{sec:SWI}
A powerful application of symmetries in quantum field theories is the
derivation of Ward and Slavnov-Taylor 
identities relating the various Green's functions and proper
vertices of a theory. Derivations
of SUSY identities have had to work around the supposed 
SUSY violating properties
of the gauge-fixing term, and the conventional approach is to replace the
SUSY parameter with ghost fields~\cite{OR02,PS86}.

We calculate the SUSY Ward identity relating the photon and photino to be
\begin{eqnarray} \label{eq:Alambda}
0 &=& \langle \delta (A_\mu(x) \lambda(y)) \rangle \nonumber \\
&=& \langle A_\mu(x) A_\beta(y) \rangle _x\partial_\alpha
\sigma^{\beta \alpha}
- \langle \lambda(y) \bar{\lambda}(x) \rangle \gamma_\mu \zeta
+ \langle \lambda(y) \bar{\lambda}(x) \rangle 
\frac{_x\partial_\mu}{_x\! \not \! \partial} \zeta,
\end{eqnarray}
where $\sigma^{\beta \alpha} 
= \frac{1}{2}(\gamma^\beta \gamma^\alpha - \gamma^\alpha \gamma^\beta)$.
Note that the last term in this equation is due to $\delta_L$, and is 
responsible for the failure of previous attempts~\cite{OR02} to derive 
this Ward identity.
Eq.~(\ref{eq:Alambda}) holds in any gauge as the $\xi$-dependant part
of the photon propagator is eliminated by multiplication with 
$\sigma^{\beta \alpha}$. In fact, eq.~(\ref{eq:Alambda}) relates the
wave renormalisation of the photino to the vacuum polarisation according to
\begin{equation} \label{eq:polarisedSUSY}
A_\lambda (p) = 1 + \Pi(p),
\end{equation} 
where the dressed photino propagator is given by
$\langle \lambda(x) \bar{\lambda}(y) \rangle =
\frac{-i}{A_\lambda (p) \not \,\! p}$ and the dressed photon propagator is
$\langle A_\mu(x) \alpha A_\beta(y) \rangle = \frac{1}{p^2}
\left(g_{\mu \nu} - \frac{p_\mu p_\nu}{p^2}\right) \frac{1}{1+\Pi(p)}
+\xi \frac{p_\mu p_\nu}{p^4}$. Indeed, eq.~(\ref{eq:polarisedSUSY}) is
what one would naively expect~\cite{KS89}.

We now investigate the identities relating the electron and selectron
propagators. It is widely believed that since their wavefunction
renormalisation is $\xi$ dependant, the SUSY violation of the gauge-fixing
term would cause the electron and selectron wavefunction renormalisations 
to differ, at least nonperturbatively. However our transformations are
not violated by covariant gauge-fixing so such reasoning does not apply.
The Ward identity relating the $\psi$ and $a$ propagators is
\begin{eqnarray} \label{eq:apsi}
0 &=& \langle \delta_{WZL} (\bar{\psi}(x) a(y)) \nonumber \\
&=& -i\bar{\zeta} \langle \psi(y) \bar{\psi}(x) \rangle
+ \bar{\zeta} \langle a(y) f^*(x) \rangle 
-i\bar{\zeta} _x\not \! \partial \langle a(y) a^*(x) \rangle,
\end{eqnarray}
as found originally by \cite{IZ74}. 
The non-linear contribution to this and
all propagator Ward identities vanishes by the cluster decomposition
principle~\cite{OR02,M93,W96}. 
This is an important result and one which should simplify
further analyses. 

\section{Discussion}
In the spirit of Wess and Zumino we have found a set of SUSY-based 
transformations that leave the action of SUSY QED completely
invariant, essentially completing their work. 
Pleasingly, these transformations obey the SUSY algebra up to the 
Lorentz gauge-fixing condition, so they obey it completely in Lorentz gauge.
While the derivation given is straightforward, the ramifications of this work
are significant. The most important consequence is of course that 
an exact SUSY transformation of Abelian gauge-field theories \textit{does} 
exist, even in Lorentz gauge. This should open
the way for much simpler analysis of these theories. In 
particular, the necessity of ghost fields for deriving SUSY identities
relating the Green's functions as claimed in some recent works~\cite{OR02} 
is seen to be false, as  is the conventional belief (\textit{eq}.~\cite{RSS01})
that only gauge invariant terms can be supersymmetric. 
Our corrected SUSY transformation can used to
derive exact Ward identities that relate the photon and photino, and
we can rederive the original identities relating the propagators of the
electron and selectrons, 
thus showing that they are not disrupted by the gauge-fixing term in spite of
the wave-function renormalisation dependance on the gauge parameter. 

The implications of this work run very deep. While 
the use of SUSY Slavnov-Taylor identities is common practice in mathematical
analyses, 
SUSY Ward identities are still a stock tool, especially in lattice field
theory~\cite{FF02a}. It seems likely that their use can be broadened
a great deal further, at least for Green's functions. 
Even if ghosts are used to linearise the SUSY transformations
or remove the auxiliary fields, current BRST transformations are based 
on the transformations 
derived by Wess and Zumino~\cite{WZ74}, which are incomplete.

While we have worked here
only in Lorentz gauge, the same general approach should be applicable in
other component form gauge choices. The extension of this work to non-Abelian
theories is very technically challenging, but necessary if this approach is to
be applied to realistic theories.

\section*{Acknowledgments}
This research was mainly funded by the 2003 Research Fund of 
Kyung Hee University, although some of the
early work was done while the author was supported by the BK21 program of
Seoul National University.

\end{document}